\documentclass{iau_FM}
\usepackage{graphicx}

\newcommand{\kms}{$\,$km$\,$s$^{-1}$}

\newcommand{\msun}{{$M_\odot$}}
\newcommand{\msunyr}{{$M_\odot$ yr$^{-1}$}}

\def\HI{H{\,\small I}}

\def\emph#1{{\sl #1}}
\newcommand{\ltsima} {$\; \buildrel < \over \sim \;$}
\newcommand{\gtsima} {$\; \buildrel > \over \sim \;$}
\newcommand{\lta} {\lower.5ex\hbox{\ltsima}}
\newcommand{\gta} {\lower.5ex\hbox{\gtsima}}

\title[Structure of  jet-driven \HI\ outflows] 
{The parsec-scale structure of jet-driven \\ \HI\ outflows in radio galaxies} 

\author[Morganti et al.]   
{Raffaella Morganti$^{1,2}$, Robert Schulz$^{1}$, Kristina Nyland$^{3}$, Zsolt Paragi$^{4}$, Tom Oosterloo$^{1,2}$,  Elizabeth Mahony$^{5,6}$, Suma Murthy$^{1,2}$
}
\affiliation{$^1$ASTRON, the Netherlands Institute for Radio Astronomy, Oude Hoogeveensedijk 4, 7991 PD Dwingeloo, The Netherlands.  email: {\tt morganti@astron.nl} \\[\affilskip] 
$^2$Kapteyn Astronomical Institute, University of Groningen, P.O. Box 800,
9700 AV Groningen, The Netherlands;  $^3$ National Radio Astronomy Observatory, Charlottesville, VA 22903, USA; \\ $^4$ Joint Institute for VLBI ERIC, Oude Hoogeveensedijk 4, 7991 PD Dwingeloo,  Netherlands; $^5$ Sydney Institute for Astronomy, School of Physics A28, The University of Sydney, NSW 2006, Australia; $^6$ ARC Centre of Excellence for All-Sky Astrophysics (CAASTRO), Australia
}

\pubyear{2018}
\jname{Astronomy in Focus, Volume 1} 
\editors{Volker Beckmann \& Claudio Ricci, eds.}
\begin{document}
\maketitle

\begin{abstract}
Radio jets can play multiple roles in the feedback loop by regulating the  accretion of the gas, by enhancing gas turbulence, and by driving gas outflows. Numerical simulations are beginning to  make detailed predictions about these processes. Using high resolution  VLBI observations we test  these predictions by studying how radio jets of different power and in different phases of evolution  affect the properties and kinematics of the surrounding  \HI\ gas.  
Consistent with  predictions, we find that young (or recently restarted) radio jets have  stronger impact  as shown by the presence of \HI\ outflows. The outflowing medium is clumpy {with clouds  of with sizes up to a few tens of pc and mass $\sim 10^4$\msun) already in the region close to the nucleus ($< 100$ pc), making the jet  interact strongly and shock the surrounding gas. We present a case of a low-power jet where, as  suggested by the simulations, the injection of energy may produce an increase in the turbulence of the medium  instead of an outflow. }
\end{abstract}


The impact of active galactic nuclei (AGN) on the surrounding medium can be due to  either winds and radiation from the nuclear region, or to plasma jets. Both these mechanisms are known to play a role and, depending on the situations and on the physical conditions, one can dominate via a strong coupling  with the surrounding medium \cite[(e.g.\  Cielo et al.\ 2018)]{Cielo18}. However, quantifying the actual impact of these phenomena is still a challenging task.  
Radio jets play a particularly important role in the feedback process, providing the best examples of AGN-driven feedback seen in action by preventing the cooling of the X-ray gas on cluster scales.  However, radio jets can also provide an effective mechanism on {\sl galaxy scales}. Their impact manifests itself in different ways: by counterbalancing the cooling of the hot coronae present around even isolated galaxies \cite[(e.g.\ Croston et al.\ 2008; Ogorzalek et al.\ 2017)]{Croston08,Ogorzalek17}; by driving fast outflows traced by different gas phases or by injecting turbulence in the ISM, e.g.\ \cite{Alatalo15,Guillard15}. All these mechanisms are relevant and need to be quantified.

Particularly interesting  are the results from recent numerical simulations \cite[(Wagner et al.\ 2012, Mukherjee et al.\ 2018a, 2018b, Cielo et al.\ 2018)]{Wagner12,Mukherjee18a,Mukherjee18b,Cielo18}.
These  show that radio jets can couple more strongly  to the ISM if that is modelled to be clumpy \cite[(Wagner et al. 2012)]{Wagner12}.  
Furthermore, a connection is expected with the cycle of activity of radio galaxies: given their small size, young (and recently restarted) radio jets have the highest impact on the gas. A dependence is also expected with  jet power:  powerful jets can drive  faster outflows while low power jets can be "trapped" for longer times and  induce more turbulence in the galactic ISM \cite[(Mukherjee et al. 2018b)]{Mukherjee18b}. Finally, the orientation of the jet with respect to the distribution of gas in the host galaxy is also relevant for the impact of the jet. 

 AGN-driven and  jet-driven outflows are known to be multi-phase and can be traced also by atomic neutral hydrogen. This has opened the possibility to test the impact of plasma jets using radio data and, in particular, 21-cm \HI\ observed in absorption (see Morganti \& Oosterloo 2018 for an overview).
The advantage of this is that the gas can be traced down to very small scales and the location of the outflow and their properties can be studied. This can be done using (sub-)arcsec down to milli-arcsec data (i.e.\ down to pc scales) as shown by the global Very Long Baseline Interferometry (VLBI) data described below and obtained by arrays including
telescopes from the European VLBI Network, the Very Long Baseline
Array (VLBA), as well as Arecibo.  
The results  allow us to investigate not only the impact of radio jets, but also whether the predictions from the simulations are confirmed.

\begin{figure}[b]
\begin{center}
 \includegraphics[angle=0,width=4.5in]{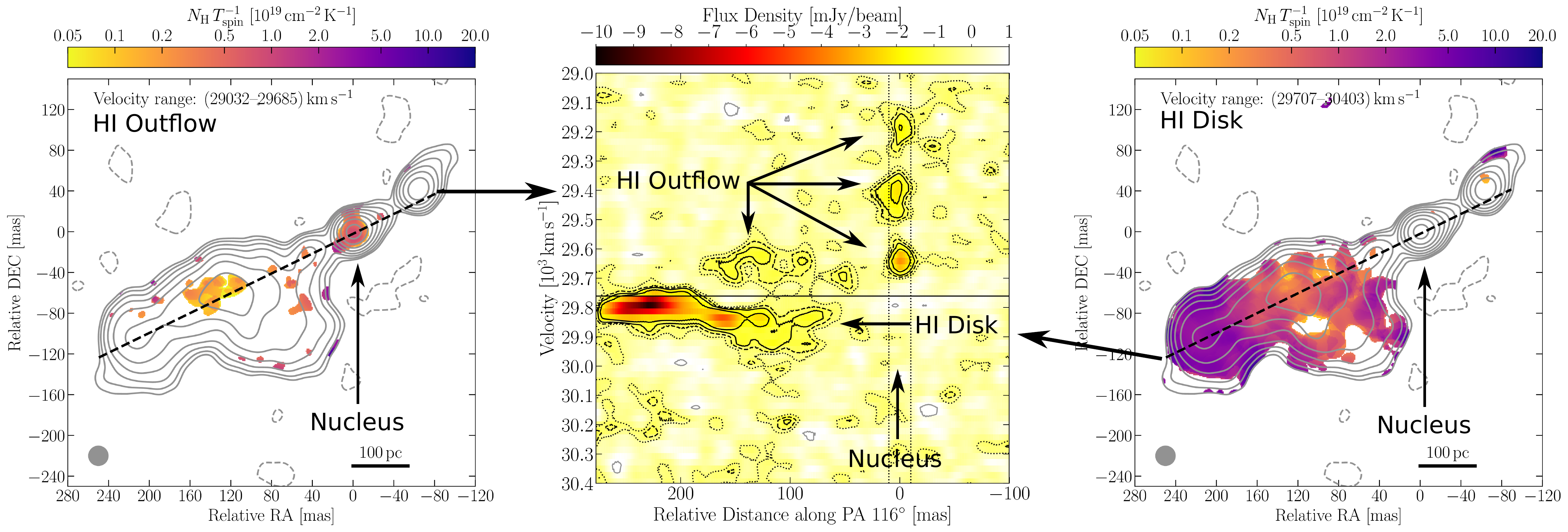} 
 \caption{Radio continuum superposed on the \HI\ absorption column density of the
\HI\ outflow (left) and of the \HI\ disk (right)  of the radio galaxy 3C236 obtained with VLBI. Position-velocity plot (centre) along the jet showing the outflowing clouds. From \cite{Schulz18}.}
   \label{fig:fig1}
\end{center}
\end{figure}

\section{Where and how often do we see  jet-driven outflows?}

Jet-driven outflows are long known from ionised gas, but more recent work has not only shown that also atomic neutral hydrogen (\HI) and molecular gas  can be associated with these outflows and also that they may carry a significant (possibly the largest) fraction  of the outflowing gas mass. The jet-driven origin of outflows of cold gas  has been confirmed in a number of cases traced by molecular gas  (see e.g.\ Alatalo et al.\ 2011, Dasyra et al.\ 2012, Combes et al.\ 2013, Morganti et al.\ 2015, Garcia-Burillo et al.\ 2014, Oosterloo et al.\ 2017, Runnoe et al.\ 2018) and  by \HI\, see \cite{Morganti18} for a review. The \HI\ outflows have typically velocities between a few hundred to $\sim$1300 \kms.

In addition to this, a relation between the occurrence of \HI\ outflows and the evolutionary status of the radio jet has been found by  observations of a relatively large sample (248 objects) presented in  \cite{Gereb15}. They find that at least 5\% of  all sources (15\% of \HI\ detections) show \HI\ outflows. These numbers represent lower limits, given that absorption measurements are sensitive  only to gas (and  outflows)  located in front of the radio continuum. Particularly interesting is that the vast majority of the \HI\ outflows are detected  in sources with newly born (or reborn) radio jets. This supports  the idea that these phases in the evolution  are those where the jet has most of its impact on the surrounding medium.   This is in agreement with   predictions from the simulations of Wagner et al.\ (2012) and Cielo et al.\ (2018).
The recurrent nature of radio sources (see e.g.\ \cite[Morganti 2017]{Morganti17}) would ensure that this impact is repeated during the life of the host galaxy.
A similar effect was also seen in the ionised gas (see Holt et al. 2008). However, this phase of the gas was found to show mass outflow rates reaching at most 1 \msunyr, while  mass outflow rates up to 50 \msunyr\ have been found for the \HI\ outflows.

\section{Do we see the interaction of the jet with a clumpy medium?}

For a small number of objects we have used VLBI observations to trace the properties of the \HI\ outflows down to pc scales. 
The results show that a clumpy distribution of the gas is seen in all observed objects. Fig.\  \ref{fig:fig1} illustrates the distribution of \HI\ absorption in the central region of the restarted radio galaxy 3C~236, \cite{Schulz18}. Interestingly, in this and other  targets observed so far, fast outflowing clouds (many hundred \kms) are detected already in the very inner region, at distances $< 50 - 100$ pc from the core. The clouds have  masses of a few $\times 10^4$ \msun, and are unresolved on VLBI scales ($< 40$ pc). 
The presence of a clumpy medium (see also Oosterloo et al.\ 2017) is of key importance and confirms the prediction of the numerical simulations. A clumpy medium can make the impact of the jet much larger than previously considered: because of the clumpiness of the medium, the jet is meandering through the ISM to find the path of minimum resistance and so creating an overpressured cocoon of outflowing and shocked gas, as suggested by \cite{Wagner12,Mukherjee18a}.

Furthermore, for the smaller (and perhaps younger) sources in the sample (4C~12.50, Morganti et al.\ 2013 and 4C~52.37, Schulz et al.\ in prep) 
the VLBI observations not only spatially resolve the outflows, but they also recover all the \HI\ flux observed at low resolution. This suggests that  these outflows are mostly made up  by relatively compact structures, easy to be detected at the very high resolution of VLBI observations. 
In the largest (and likely more evolved) sources, like 3C~293 (Schulz et al. in prep) and 3C~236 (Schulz et al. 2018),  we also find evidence of a clumpy structure but
the \HI\ outflows in these sources are only partly recovered by VLBI. This suggests the additional presence of a diffuse component in which the clumps are embedded. This could be due to the expansion of the jet in the medium changing the structure of the outflows, the fraction of diffuse component increasing with time.

\begin{figure}
\begin{center}
 \includegraphics[angle=0,width=4in]{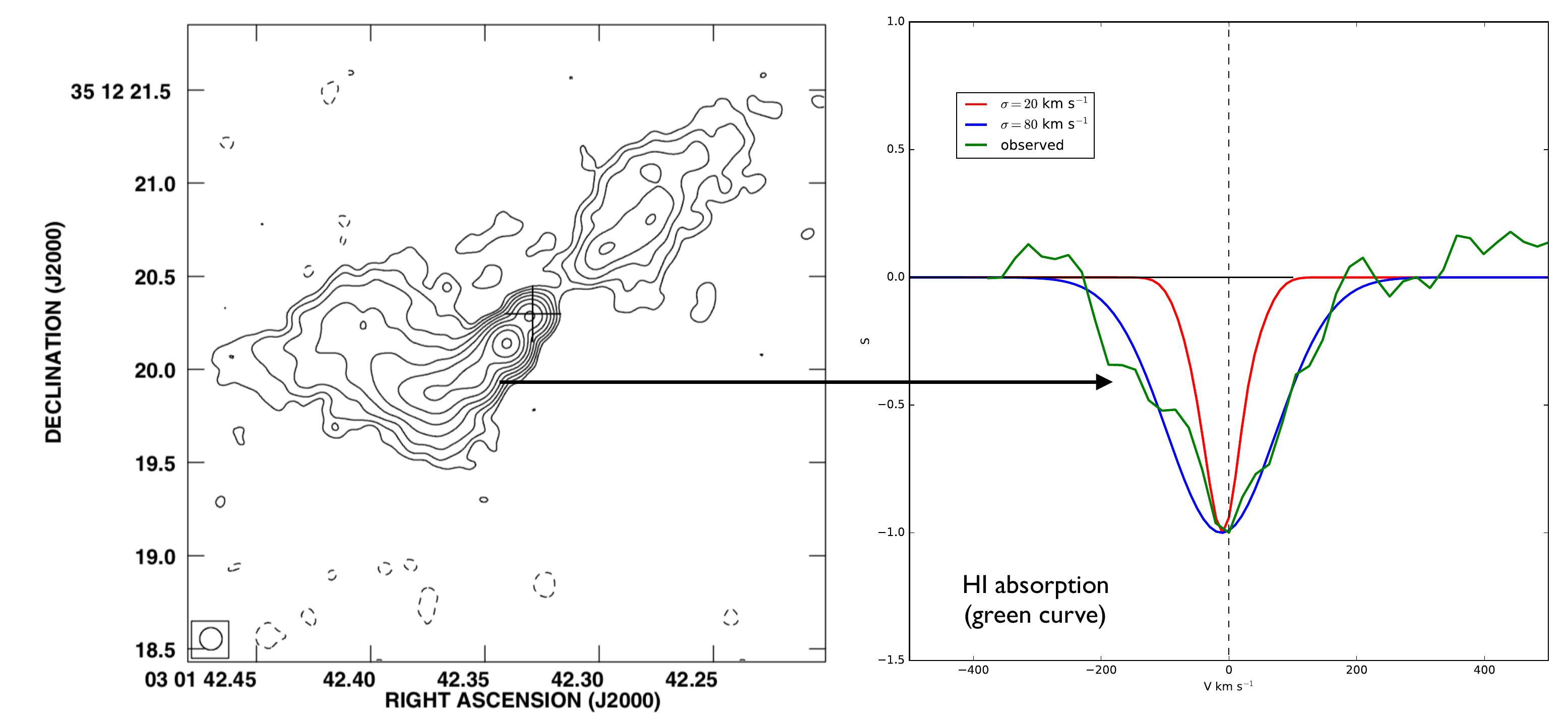}  
 \caption{{\bf Left:} Radio continuum image from \cite{Giroletti05}. {\bf Right:} \HI\ absorption profile (green) from VLA observations with superposed (red) the model from the \HI\ disc observed in emission in this object by \cite{Struve10} and in blue the model adding a component of turbulence due to the interaction of the jet with the \HI\ in the disk  (Murthy et al.\ in prep).}
   \label{fig:fig2}
\end{center}
\end{figure}

\section{Do the low power jets have also an impact?}

Interestingly, a growing number of cases (among which some listed above e.g.\ NGC1433, IC~5063, NGC1068, PG1700+518) show that, despite being classified as {\sl radio quiet}, the power of the jet is sufficient to be the driving mechanism of their outflows. 
However, the simulations also show that an other effect expected from low power jets (Mukherjee et al.\ 2018b) is  to  increase the turbulence of the gas. Figure \ref{fig:fig2} shows the \HI\ absorption detected against the kpc-scale jet of the low-luminosity radio source  B2~0258+35. The width of the absorption ($\sim 400$ \kms) is too large to be explained by the rotation  of the large scale \HI\ disk known to be present in this object.  The most likely hypothesis is that the jet enters the disk and, being trapped there, disturbs the kinematics of the gas without being able to produce a fast outflow, but injects energy  increases the turbulence of the gas (Murthy et al.\ in prep). As already suggested for the low-power radio source NGC~1266 (Alatalo et al. 2015),  jet-induced turbulence may play a role in preventing star formation  despite the large reservoir of cold gas observed in these objects. 

\section{Implications}
The observations are showing  evidence - in a growing number of sources - of interaction between the radio jets and the surrounding ISM. The properties appear to be, to first order, consistent with the predictions from some of the recent numerical simulations. This supports the idea that also on galactic scales the role of radio jets  should not be neglected. However, the impact of outflows may not always be as large as required by models of galaxy evolution.  A relatively small fraction of the outflowing gas may actually leave the galaxy. This is also seen in more AGN-driven outflows, i.e.\  those driven by winds or radiation. 
Thus, the likely main effect of jet-ISM interactions and their injection of energy could be on the redistributing the gas and possibly in keeping it turbulent for longer periods of time.  This has been now seen in particular for (much more common) low luminosity radio jets. Thus, in addition to the search for violent processes like outflows, other more subtle effects needs to be searched for and investigated.

\end{document}